\documentclass[
 reprint,
bibnotes,
 amsmath,amssymb,
 aps,
dvipdfmx,
twocolumn
]{revtex4-2}

\usepackage{graphicx}
\usepackage{dcolumn}
\usepackage{bm}
\usepackage{braket}
\usepackage{cases}
\usepackage{amssymb}
\usepackage{color}
\usepackage{url}
\usepackage{algorithm}
\usepackage{algpseudocode}

\begin{document}

\title{Continuous percolation in a Hilbert space for a large system of qubits}
 
\author{Shohei Watabe$^{1,2,3}$, Michael Zach Serikow$^{4,5}$, Shiro Kawabata$^6$, and Alexandre Zagoskin$^{7}$}  
\affiliation{$^1$ Division of Nano-quantum Information Science and Technology, Research Institute for Science and Technology, Tokyo University of Science, Shinjuku, Tokyo 162-8601, Japan}
\affiliation{$^2$ College of Engineering, Department of Computer Science and Engineering, Shibaura Institute of Technology, 3-7-5 Toyosu, Koto-ku, Tokyo 135-8548, Japan}
\affiliation{$^3$ Division of Nano-quantum Information Science and Technology, Research Institute for Science and Technology, Tokyo University of Science, Shinjuku, Tokyo 162-8601, Japan}
\affiliation{$^4$ Department of Physics, University of Notre Dame, Notre Dame, IN 46556, USA} 
\affiliation{$^5$ Notre Dame Institute for Advanced Study, University of Notre Dame, IN 46556, USA} 
\affiliation{$^6$ Research Center for Emerging Computing Technologies (RCECT), National Institute of Advanced Industrial Science and Technology (AIST),
1-1-1 Umezono, Tsukuba, Ibaraki 305-8568, Japan,}
\affiliation{$^7$ Department of Physics, Loughborough University, Loughborough LE11 3TU, UK}

\begin{abstract} 
The development of percolation theory was historically shaped by its numerous applications in various branches of science, in particular in statistical physics, and was mainly constrained to the case of Euclidean spaces. One of its central concepts, the percolation transition, is defined through the appearance of the infinite cluster, and therefore cannot be used in compact spaces, such as the Hilbert space of an $N$-qubit system. Here we propose its generalization for the case of a random space covering by hyperspheres, introducing the concept of a ``maximal cluster". Our numerical calculations reproduce the standard power-law relation between the hypersphere radius and the cover density, but show that as the number of qubits increases, the exponent quickly vanishes (i.e., the exponentially increasing dimensionality of the Hilbert space makes its covering by finite-size hyperspheres inefficient). 
Therefore the percolation transition is not an efficient model for the behavior of multiqubit systems, compared to the random walk model in the Hilbert space. However, our approach to the percolation transition in compact metric spaces may prove useful for its rigorous treatment in other contexts.

\end{abstract}
\maketitle


\section{introduction}

As much of mathematics, the development of the percolation theory was initiated and stimulated by the needs of physics (in this case, the transport phenomena) and engineering. The continuum percolation in particular is of much interest because of the wide range of applications, and it was extensively investigated in a number of situations (on plane, torus, Klein bottle and their higher-dimensional analogues~\cite{stauffer1992,Mertens2012} as well as on the projective plane~\cite{Freedman1997,Borman2016}). 

The question of accessibility of different regions of the Hilbert space is especially relevant for quantum information processing, in particular, quantum annealing and adiabatic quantum computing~\cite{Kadowaki1998,Albash2018}. 
Quantum annealing is based on the adiabatic theorem, 
where an easily achieved, factorized initial ground state slowly transforms into the final factorized by design ground state that provides a solution to a combinatorial optimization problem~\cite{Kadowaki1998,Albash2018}. 
In a system of $N$-qubits, the overlap between these states scales as $1/2^N$. 
Therefore, the quantum state of a quantum annealer traces a path between two almost orthogonal factorized states.

A single quantum trajectory in the presence of decoherence can be approximated by a series of consecutive unitary evolutions governed by the time-dependent Hamiltonian $H(t)$~\cite{Percival2008}, which are interrupted (on average) every $t_D$ by a collapse (a projective measurement of some relevant operator) fixing the system in some pure, generally not globally entangled state. The distance covered during each unitary evolution is restricted by the Margolus-Levitin theorem and its generalizations~\cite{Margolus1998,Deffner2013,Okuyama_2018}, such as
\begin{equation}
\tau_{if} \geq \frac{\hbar}{2\bar{E}} \sin^2 \Theta_{if},
\label{eq:Deffner2013}
\end{equation} 
where $\tau_{if}$ is the time necessary to evolve between pure states $|\psi_i\rangle$ and $|\psi_f\rangle$ under the action of the Hamiltonian $H(t)$, the average energy $\bar{E} = \tau_{if}^{-1} \int_0^{\tau_{if}} \langle H(t)\rangle dt$, and the Bures angle $\Theta_{if} = \arccos[ |\langle\psi_i|\psi_f\rangle | ].$ 
Note that this angle provides a unique nontrivial parametrization of the Hilbert space (Fubini-Study metric).
In the above approximate picture, a quantum trajectory will lie within a union of hyperspheres of radius $R_D = R_D (t_D)$.

The stochastic quantum trajectory model is a numerically efficient approach towards the modeling of quantum systems with dissipation~\cite{wiseman2010quantum}. The quantum state diffusion~\cite{Percival2008} employs the Brownian motion limit for quantum trajectories. An approximation of a general random quantum walk was useful in evaluating the likelihood of success of quantum annealing and deriving its sufficient criterion, the quantum accessibility index~\cite{Watabe2022}, which is a numerical characteristic of the likeliness of an adiabatic transition between two orthogonal quantum states. 

The realization of the fundamental impossibility of an efficient simulation of large enough quantum coherent systems with classical means~\cite{Feynman1982,Manin1980} led to the development of quantum technologies, which demonstrated spectacular results in a rather short time~\cite{QuantumManifesto2016}. 
As the result, such large enough quantum devices have been developed~\cite{DWave2000X,Bunyk2014,Walport2016,Arute2019,IBM127,Gong2021,Ebadi2021,Madsen2022,arxiv.2209.06841}. 
This made explicit the necessity of developing model-independent, qualitative approaches towards the description of quantum systems with huge Hilbert spaces, such as scaling theories.

Since the random walk and percolation are closely related~\cite{stauffer2018introduction}, such considerations naturally lead to the attempt to the description of the evolution of a quantum system in terms of a percolation problem in the Hilbert space. 
In particular, it was very tempting to consider whether a finite probability of success of an adiabatic quantum computer arises due to the percolation transition in its Hilbert space. The simplest model would be a continuous percolation through randomly placed hyperspheres. Due to compactness, and actually finite diameter of the Hilbert space~\cite{Brody2001}, the standard definition of a percolation transition (appearance of the infinite cluster~\cite{stauffer1992}) does not apply. 

Therefore, we generalize the percolation transition problem to the case of a compact metric space, and we replace the infinite cluster with the ``maximal span cluster", MSC, that is, a cluster which contains at least two points separated by the maximal Fubini--Study distance. 
One expects that the minimal number $M$ of hyperspheres in a cover, which would contain at least one MSC, should scale as a negative power of the $R_D$. As will be shown later, this was confirmed by the numerical calculations with the form 
$ M \sim R_D^{-a D^{b}}$,
where $D$ is the dimensionality of the Hilbert space (for $N$ qubits, $D=2^N$) and fitting parameters $a,b(>0)$. This result, unfortunately, means that the number $M$, and therefore the conditions of the transition, very quickly become practically independent on $R_D$, and cannot be used to characterize the accessibility of the Hilbert space, unlike the random walk model~\cite{Watabe2022}. 
We also prove the concentration inequality between two almost orthogonal random quantum states. This inequality may be useful for testing whether randomly generated quantum states are uniformly distributed in the Hilbert space. 
So, while the results of our approach proved useless for its initial purpose, they may prove useful for the theory of percolation in compact spaces.

\section{Percolation in the Hilbert Space}

\begin{figure}[tbp]
      \centering
      \includegraphics[width=80mm]{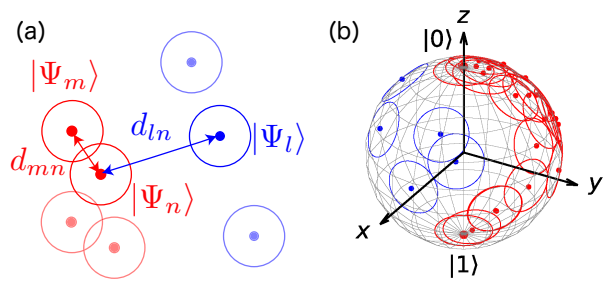}
      \caption{
      Percolation in the Hilbert space. 
      (a) Schematics of connection between states. The points represent states generated randomly, and the diameter of the circle shows the distance $\Delta S$. Red/blue shows connected/disconnected states. 
      (b)  Schematic of the percolation on the Bloch sphere. 
      The red/blue points indicate the random states generated uniformly on the Bloch sphere, 
      where the red/blue points can/cannot be organized into a cluster that maximally spans the Bloch sphere. The diameter of the red and blue circles is $\Delta S$. The red cluster spans from the state $| 0 \rangle$ to the orthogonal state $| 1 \rangle$. 
      }
      \label{Fig_SchematicPercolation_1}
\end{figure}

We consider a percolation model of adiabatic state evolution in a partially quantum coherent system. The simplest case would be a continuous percolation through randomly placed hyperspheres. We here introduce the notion of ``maximal span cluster", MSC, that is, a cluster which contains at least two points separated by the maximal Bures angle, $\pi$, instead of the standard percolation transition model with the infinite cluster~\cite{stauffer1992}) that cannot not be applied in the present case because of the compactness and finite diameter of the Hilbert space~\cite{Brody2001}.

We formulate the continuum percolation transition in a finite-dimensional Hilbert space randomly covered by hyperspheres (see also Algorithm~\ref{algo1}). 
First, $M$-states $|\Psi_n \rangle$ are randomly prepared for $n = 1, 2, \cdots M$, which are uniformly distributed in the Hilbert space. 
Clusters are then created by connecting two states $|\Psi_{m} \rangle$ and $| \Psi_n \rangle$ with the condition where the Fubini--Study distance $d_{mn} \equiv s (\Psi_m, \Psi_n) = \cos^{-1} (|\langle \Psi_m | \Psi_n \rangle|)$ is shorter than a certain threshold $\Delta S$, {\it i.e.}, $d_{mn} = s (\Psi_m, \Psi_n) \leq \Delta S$ (see Fig.~\ref{Fig_SchematicPercolation_1} (a)). 
Let $\alpha$ be the label of a cluster. By comparing the distances between all the states in the $\alpha$-th cluster, we can find the maximum distance $L_\alpha =  {\rm max} (s (\Psi_m, \Psi_n))$, where $\Psi_{m,n}$ are the states that belong to the $\alpha$-th cluster. 
If we find $| \pi/2 - L_\alpha | \leq \epsilon$ for some small $\epsilon$, we regard that the $\alpha$-th cluster spans maximally in the Hilbert space (see Fig.~\ref{Fig_SchematicPercolation_1} (b)).

\begin{figure}[!t]
\begin{algorithm}[H]
  \caption{Percolation in Hilbert space}
  \label{algo1}
   \begin{algorithmic}[1]
   \State Uniformly generate at random $|\Psi_n \rangle$  for $n = 1, 2, \cdots M$ by using the normal distribution. 
   \For{$m,n = 1 \, \ldots \, M$}
      \State{calculate the distance $s(\Psi_m,\Psi_n)$}
   \EndFor
   \For{$m,n = 1 \, \ldots \, M$}
      \If{$s(\Psi_m,\Psi_n) < \Delta S$} 
          \State make a cluster by connecting $m$ and $n$. 
      \EndIf 
   \EndFor
   \State Find $L_\alpha = {\rm max}(s(\Psi_m,\Psi_n))$ in each cluster, where $m,n$ belong the $\alpha$-th cluster.
   \If{$| \pi/2 - L_\alpha | \leq \epsilon$} 
       \State Judge that the cluster spans maximally in the Hilbert space. 
   \EndIf 
   \end{algorithmic}
\end{algorithm}
\label{FigAlgo1}
\end{figure}

\begin{figure}[!t]
\begin{algorithm}[H]
  \caption{Creation of Boolean List for cluster}
  \label{algo2}
   \begin{algorithmic}[1]
      \State Initially $b_{mn} = {\tt F}$
      \For{$m,n = 1 \, \ldots \, M$}
            \If{$s(\Psi_m,\Psi_n) < \Delta S$} 
                \State $b_{mn} = {\tt T}$ 
                  \For{$l = 1 \, \ldots \, M$}
                        \State $b_{ml}, b_{nl}= b_{ml} \lor b_{nl}$
                  \EndFor
            \EndIf 
      \EndFor

      \For{$l,m,n = 1 \, \ldots \, M$ where $m > n$}
            \If{$b_{ml} = b_{nl}$} 
                \State $b_{ml}$ with ${\tt F}$
            \EndIf 
      \EndFor

      \State Initialize $\alpha = 0$
      \For{$n = 1 \, \ldots \, M$}
      \For{$m = 1 \, \ldots \, M$}
      \If{$b_{mn} = {\tt T}$} 
          \State Make a cluster list $c_{\alpha} = \{ b_{1n}, b_{2n}, \cdots , b_{Mn}\}$ 
      \EndIf 
      \EndFor
      \State $\alpha = \alpha + 1$
      \EndFor
   \end{algorithmic}
\end{algorithm}
\end{figure}

In the numerical simulation, we take the following procedures.  
We first randomly prepare $M$-complex vectors $|\Psi_n \rangle = (u_{1}^{(n)} , u_{2}^{(n)}, \cdots,  u_{D}^{(n)} )^{\rm T}$  for $n = 1, 2, \cdots M$, where $D$ is the dimension of the Hilbert space related to the number of qubits $N$ as $D = 2^{N}$. 
For uniformly preparing normalized states at random in the Hilbert space, we generate $2D$-real values $(x_1, x_2, \cdots, x_{2D})$ from the normal distribution ${\mathcal N}(0, 1)$, and provide a complex number $u_j^{(n)} = (x_{2j-1} + i x_{2j}) / r$ for $j = 1, 2, \cdots,  D$, where $r = \sqrt{x_1^2 + x_2^2 + \cdots + x_{2D}^2}$. 
We then evaluate the Fubini--Study distance $d_{mn} = s(\Psi_m,\Psi_n) = \cos^{-1} (|\langle \Psi_m | \Psi_n \rangle|)$ between all pairs of random states for $m,n = 1, 2, \cdots , M$.

We then construct clusters of states that are linked with the condition $s(\Psi_m,\Psi_n) < \Delta S$. This can be executed in the following way: 
First, we prepare an initial boolean array $b_{mn} = {\tt F}$(false) for all pairs of states. (For the treatment of the Boolean list for the cluster, see also Algorithm~\ref{algo2}). 
We then replace $b_{mn} = {\tt F}$ with ${\tt T}$(True) if $s(\Psi_m,\Psi_n) \leq \Delta S$ for $m,n = 1, 2, \cdots , M$. 
If states $|\Psi_{m} \rangle$ and $|\Psi_n\rangle$ are connected, i.e., $b_{mn} = {\tt T}$, we make $b_{ml}, b_{nl}= b_{ml} \lor b_{nl}$ for $l = 1,\cdots, M$, where $b_{ml}$ and $b_{nl}$ share ${\tt T}$, i.e., information of states connected to the states $m$ and $n$. 
The boolean array $b_{mn}$ has then all information of random states that belongs to a cluster. 

Let $N_{\rm c}$ be the number of clusters generated by the rule shown above. 
In order to minimize the duplication of cluster information, we replace $b_{ml}$ with ${\tt F}$ if $b_{ml} = b_{nl}$ for $m>n$. 
We then generate the boolean array for the cluster $c_{\alpha} = \{ b_{1n}, b_{2n}, \cdots , b_{Mn}\}$ such that at least one $b_{mn}$ is ${\tt T}$ for $m = 1, \cdots, M$. Here, the index $\alpha$ is the label of the cluster, the number of which is $N_{\rm c}$.  
If $b_{ln} = b_{mn} = {\rm T}$ for $b_{ln,mn} \in c_\alpha$, the states $|\Psi_l\rangle$ and $|\Psi_m \rangle$ belong to the $\alpha$-th cluster, the distance of which is $s(\Psi_l,\Psi_m)$. 
By comparing all the distances between states in the $\alpha$-th cluster, we can find the maximum size $L_\alpha$ of the $\alpha$-th cluster. If it satisfies $\pi/2 - L_\alpha \leq \epsilon$, we judge the $\alpha$-th cluster to be span maximally in the Hilbert space.  

\begin{figure}[tbp]
      \centering
      \includegraphics[width=80mm]{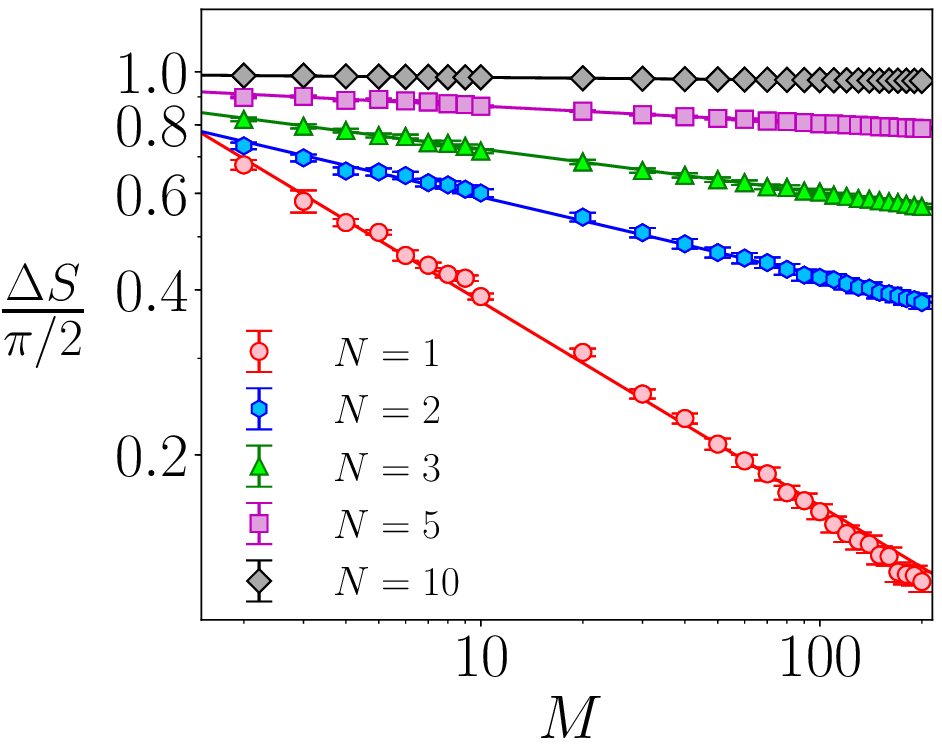}
      \caption{
      The Fubini-Study threshold distance $\Delta S$ for connecting two states as a function of the number of isotropic random states $M$ in the Hilbert space for $N$-qubits, where the cluster can expand the maximum Fubini-Study distance $\pi/2$. The data points are the averaged value for $100$ samples, and solid lines are fitted lines with $\Delta S = (\pi/2)A M^{-B}$ for $M = 2$--$200$.
      }
      \label{Fig_SchematicPercolation_2}
\end{figure}

\begin{figure}[tbp]
      \centering
      \includegraphics[width=80mm]{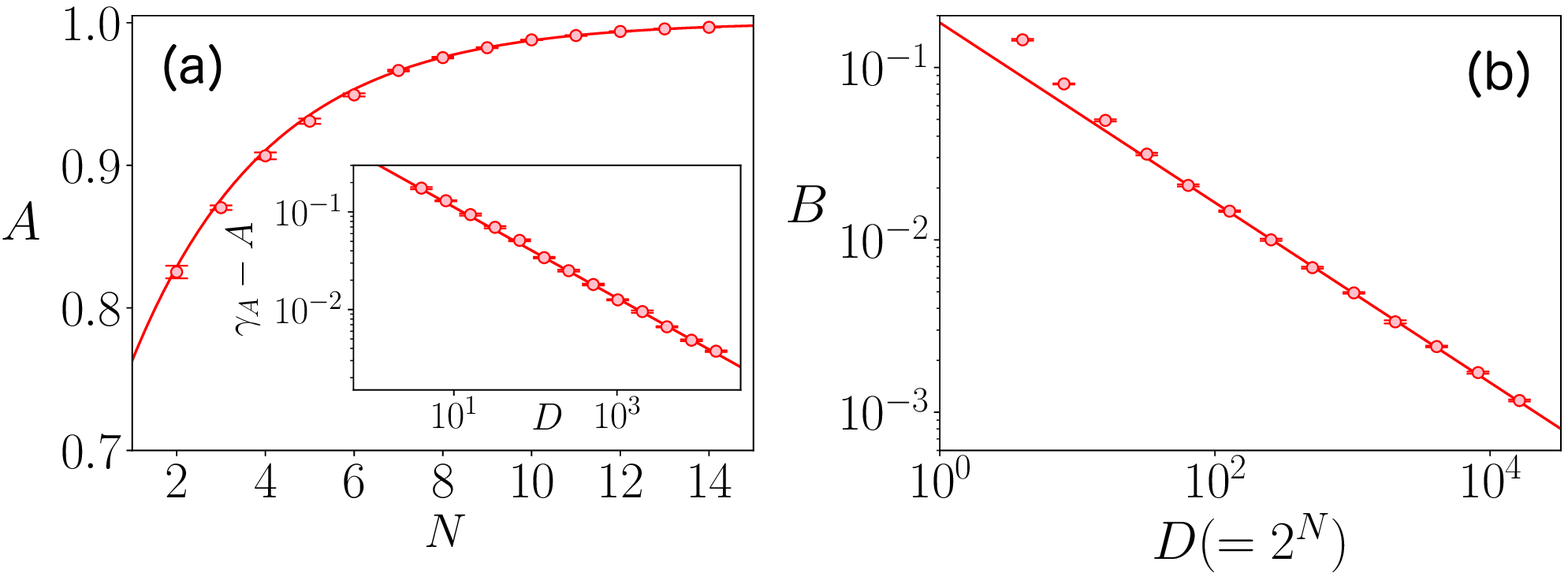}
      \caption{
      (a) The fitting factor $A$ and (b) the fitting factor $B$ as a function of the number of qubits $N$ and the dimension of the Hilbert space: $D = 2^N$. 
      The solid lines represent fitting functions, where we used $A = \gamma_A - \alpha_A D^{-\beta_A}$ and $B = \alpha_B D^{-\beta_B}$ for $N=7$--$14$.  The inset in (a) is a log-log plot of $\gamma_A - A$ as a function of $D$. 
      }\label{Fig_SchematicPercolation_3}
\end{figure}

Figure~\ref{Fig_SchematicPercolation_2} plots the critical Fubini-Study distance $\Delta S$ as a function of the number of the random states $M$ generated uniformly in the Hilbert space. 
Here, the critical Fubini--Study distance is determined by the threshold of the distance $\Delta S$, where a cluster starts to span maximally with the condition $\pi/2 - L_\alpha \leq \epsilon$. 
In this numerical simulation, we take $\epsilon = \Delta S$. 
The scaling of the critical Fubini-Study distance $\Delta S$ can be given by $\Delta S = (\pi/2) A M^{-B}$, where $A(D) = 1.0005(5) - 0.33(2) D^{-0.47(2)}$ and $B(D) = 0.182(4) D^{-0.522(4)}$, with the dimension of the Hilbert space $D=2^N$ (see Fig.~\ref{Fig_SchematicPercolation_3}).

A numerical simulation shows that the exponent scales to zero as a power of the dimension of the Hilbert space, so the threshold quickly becomes insensitive to the number of qubits and rules out the percolation transition as a candidate for the adiabatic accessibility index. In a system of many qubits, 
we need almost the maximum value of the distance $\Delta S \simeq \pi/2$ to make the cluster span maximal irrespective of the number of random states $M$. This property is consistent with the almost orthogonality of independent random vectors in the high-dimensions for ${\mathbb R}^n$~\cite{vershynin_2018}. 
Note that the random walk model shows the same scaling form $\Delta S = (\pi/2) A M^{-B}$, although it gives $A = 0.309 (8) N^{0.76(2)}$ and $B(N) = 0.4727(8) N^{0.020(1)}$~\cite{Watabe2022}. The dependence of $N$ on $B$ in the random walk is much smaller than that of the percolation model. 

\section{Concentration Inequality}


In the previous section, we found that the percolation transition is ruled out as a useful model of evolution of large, partially coherent quantum systems, because two uniformly generated random states $| \Psi \rangle$ and $| \Psi' \rangle$ become almost orthogonal as the dimensionality of the Hilbert space grows. 
We can show the concentration inequality for the absolute square of the inner product between these two states, given in the form 
\begin{align}
P \left (  \left | \langle \Psi | \Psi' \rangle |^2  - \frac{1}{D} \right | \geq  \epsilon \right ) \leq 4  \exp \left [ - \frac{D}{4} \left ( \frac{D \epsilon}{1+D \epsilon }\right )^2 \right ] , 
\label{ConcentrateInequality}
\end{align} 
where $  | \langle  \Psi | \Psi' \rangle |^2 $ is almost the inverse of the dimension $D = 2^{N}$, {\it i.e.} exponentially small for the number of qubits $| \langle \Psi | \Psi' \rangle |^2 = {\mathcal O} (2^{-N})$. 
(The derivation of this concentrate inequality is shown in Appendix~\ref{AppendixConcentrateInequality}.)
Taking $\epsilon = 1/D$, 
we have $P_{\rm L.B.}  \leq P \left ( 0 <  | \langle \Psi | \Psi' \rangle |^2 < 2 /2^N \right ) \leq 1 $, where  $P_{\rm L.B.} \equiv 1- 4 \exp \left ( - 2^N/16 \right )$. 
For example, the lower bound $P_{\rm L.B.}$ is $0.92\cdots$ for $6$-qubits, $0.998\cdots$ for $7$-qubits, and $0.9999995\cdots$ for $8$-qubits. 
The probability where $0 < | \langle \Psi | \Psi' \rangle |^2 < 1/2^{N-1}$ approaches to unity with the number of qubits $N$ double-exponentially. 

We naturally expect that the fidelity between two random states generated isotropically is exponentially small with respect to the number of qubits $N$: 
\begin{align}
| \langle \phi | \Psi \rangle |^2 = {\mathcal O} \left ( \frac{1}{2^N}  \right ) . 
\end{align} 
Let $| \Psi \rangle = (\Psi_1, \cdots, \Psi_D)^{\rm T}$ and $| \phi \rangle = (\phi_1, \cdots, \phi_D)^{\rm T}$ be $D$-dimensional random normalized complex vectors. 
Here, $\Psi_i$ and $\phi_i$ for $i = 1,\cdots, D$ are generated from mutually independent  random variables $u_j, v_j \in {\mathbb R}$ for $j = 1, \cdots, 2D$, where $\Psi_j = (u_{2j-1} + i u_{2j})$ and $\phi_j = (v_{2j-1} + i v_{2j})$ with $\sum\limits_{j = 1}^{2D} u_j^2 = \sum\limits_{j = 1}^{2D} v_j^2= 1$. 
The fidelity between two states is given by 
$| \langle \phi | \Psi \rangle |^2 =  ({\bf v} \cdot {\bf u})^2 + (  {\bf v}' \cdot {\bf u} )^2$, 
where ${\bf u} = (u_1, u_2, \cdots, u_{2D})^{\rm T}$, 
${\bf v} = (v_1, v_2, \cdots, v_{2D})^{\rm T}$ and ${\bf v}' = (-v_2, v_1, -v_4, v_3, \cdots, - v_{2D}, v_{2D-1})^{\rm T}$. 
If ${\bf u}$ and ${\bf v}$ are $2D$ isotropically random unit vectors, the angular distance is known to show the relation $\cos^{-1} (|{\bf v} \cdot {\bf u}| ) = \pi/2 + {\mathcal O} (1/\sqrt{2D})$~\cite{Hall2005}. 
As a result, in integrated qubit systems, two random states generated isotropically are almost orthogonal. 
The concentrate inequality~\eqref{ConcentrateInequality} will be helpful to check whether generated quantum states are uniformly random in the Hilbert space.

\section{Conclusion}

Inspired by the random walk model of the stochastic process in the quantum system, 
we have investigated a continuous percolation in the Hilbert space for studying a large system of qubits. 
The Hilbert space of a large system of qubits has a huge dimensionality, but due to its being compact the standard concepts of percolation, such as an infinite cluster, cannot be applied.
In this paper, we introduced the notion of ``maximal span cluster", where the cluster size is measured by the Fubini--Study distance between states in a cluster, and the size of the maximal span cluster size is bounded by $\pi/2$, the maximum distance between two states in the Hilbert space. 

We find that the critical Fubini-Study distance between two states, where the cluster starts to span maximally in the Hilbert space, scales as a power of the number of randomly generated states. However, the exponent scales to zero as a power of the dimension of the Hilbert space, which is stark contrast to the random walk model. This means that the states randomly generated in the Hilbert space are almost orthogonal, and the percolation model is ruled out as the accessibility index, which is the index for the quantumness of the adiabatic evolution in the large system of qubits. 
We have also derived the concentration inequality, which can quantitatively evaluate the almost orthogonality in randomly generated states. 

In conclusion, we have seen that percolation-based models cannot be the basis for an efficient characterization of adiabatic evolution in partially quantum coherent systems. Nevertheless, the mathematical problem of the percolation transition in compact metric spaces is new and interesting, and our approach may prove useful for its rigorous treatment.     

\begin{acknowledgments}
This paper is partly based on results obtained from a project, JPNP16007, commissioned by the New Energy and Industrial Technology Development Organization (NEDO), Japan. S.W. was supported by Nanotech CUPAL, National Institute of Advanced Industrial Science and Technology (AIST) and JST, PRESTO Grant Number JPMJPR211A, Japan. A.Z. was supported by NDIAS Residential Fellowship. 
\end{acknowledgments}

\appendix 

\section{Concentrate Inequality}\label{AppendixConcentrateInequality}

This appendix shows the derivation of the inequality \eqref{ConcentrateInequality}. 
We consider the concentration inequality for the fidelity between a basis $| i \rangle$ and an isotropically random states $| \phi \rangle = \sum\limits_{i=1}^D \phi_i | i \rangle$ in a $D$ Hilbert space with $\phi_j = (v_{2j-1} + i v_{2j})$ for ${\bf v} = (v_{1}, \cdots, v_{2D})^{\rm T} \in {\mathbb R}^{2D}$. 
The vector ${\bf v}$ is a $2D$ isotropically random vector, the components of which are mutually independent and can be generated from the $2D$ normal distribution ${\boldsymbol \xi} = (\xi_1, \cdots, \xi_{2D})^{\rm T} \sim {\mathcal N}_{2D} ({\bf 0}, {\bf 1})$ with a relation $v_i = \xi_i / |{\boldsymbol \xi}|$ and $| {\boldsymbol \xi} |^2 = \sum\limits_{i}^{2D} \xi_i^2 $. 
In this case, the fidelity is given by 
\begin{align} 
| \langle i | \phi \rangle |^2 = v_{2i-1}^2 + v_{2i}^2 = \frac{\xi_{2i-1}^2 + \xi_{2i}^2}{ | {\boldsymbol \xi} |^2 }. 
\end{align} 
We here consider the following probability 
\begin{align}
 P \left (  \left | | \langle i | \phi \rangle |^2  - \frac{1}{D} \right | \geq  \epsilon \right ). 
\end{align}  
This probability can be given by 
\begin{align}
& P \left (  \left | | \langle i | \phi \rangle |^2  - \frac{1}{D} \right | \geq  \epsilon \right )
 \\
 \leq & 
  P \left (   | \langle i | \phi \rangle |^2 \geq A_+ \right ) +   P \left (   | \langle i | \phi \rangle |^2 \leq A_- \right ), 
  \label{eqP+P-}
\end{align}  
where $A_\pm = 1/D \pm \epsilon$. 

By using the Fr\'echet inequality, the first term can be reduced to 
\begin{align}
P_+ \equiv 
& 
P \left (   | \langle i | \phi \rangle |^2 \geq A_+ \right )  
\\
= & 
P ( \xi_{2i-1}^2 + \xi_{2i}^2 \geq | {\boldsymbol \xi} |^2 A_+) 
\\ 
\leq & 
{\rm min} \{ P ( \xi_{2i-1}^2 + \xi_{2i}^2 \geq 2 D A_+ \eta_+ ) ,  P (  2 D \eta_+ \geq  | {\boldsymbol \xi} |^2  ) \}, 
\end{align} 
for $\eta_+ > 0$. 

The first part $P ( \xi_{2i-1}^2 + \xi_{2i}^2 \geq 2 D A_+ \eta_+ )$ can be given by 
\begin{align}
& 
P ( \xi_{2i-1}^2 + \xi_{2i}^2 \geq 2 D A_+ \eta_+ ) 
\\ 
= & P \left ( \frac{\xi_{2i-1}^2 + \xi_{2i}^2}{2}  -1 \geq  D A_+ \eta_+ -1 \right )
\\ 
\leq & P \left ( \left | \frac{\xi_{2i-1}^2 + \xi_{2i}^2  }{2} - 1 \right | \geq  D A_+ \eta_+ -1 \right ). 
\end{align} 
It is known that if $Z_k$ is generated from the normal distribution ${\mathcal N} (0,1)$, 
the concentration of $\chi^2$-variables holds: 
\begin{align}
P \left (\left | \frac{1}{n} \sum_{k=1}^n Z_k^2 - 1 \right | \geq t \right ) \leq 2 \exp \left ( - \frac{nt^2}{8} \right ), 
\label{ConcentrationChi}
\end{align} 
for $t \in (0, 1)$~\cite{wainwright_2019}. 
Using this concentration inequality, we have a tail bound: 
\begin{align}
P ( \xi_{2i-1}^2 + \xi_{2i}^2 \geq 2 D A_+ \eta_+ ) 
\leq & 2 \exp \left [ - \frac{( D A_+ \eta_+ -1)^2 }{4} \right ], 
\end{align} 
for $1/(1+D\epsilon) < \eta_+ < 2/(1+D\epsilon) $.

By using the tail bound (\ref{ConcentrationChi}), 
the second part $P (  2 D \eta_+ \geq  | {\boldsymbol \xi} |^2  )$ can be given by 
\begin{align}
P(  2D \eta_+   \geq  | {\boldsymbol \xi} |^2 )  
= & 
P( 2 D - | {\boldsymbol \xi} |^2  \geq  2D  (1 - \eta_+)   ) 
\\
\leq & 
P\left ( \left  | \frac{| {\boldsymbol \xi} |^2}{2D}   -   1 \right |  \geq  1 -\eta_+    \right  )  
\\
\leq & 2 \exp \left [ - \frac{D (1-\eta_+)^2}{4} \right ], 
\end{align} 
where $0 < \eta_+ < 1$.

\begin{figure*}[tbp]
      \centering
      \includegraphics[width=140mm]{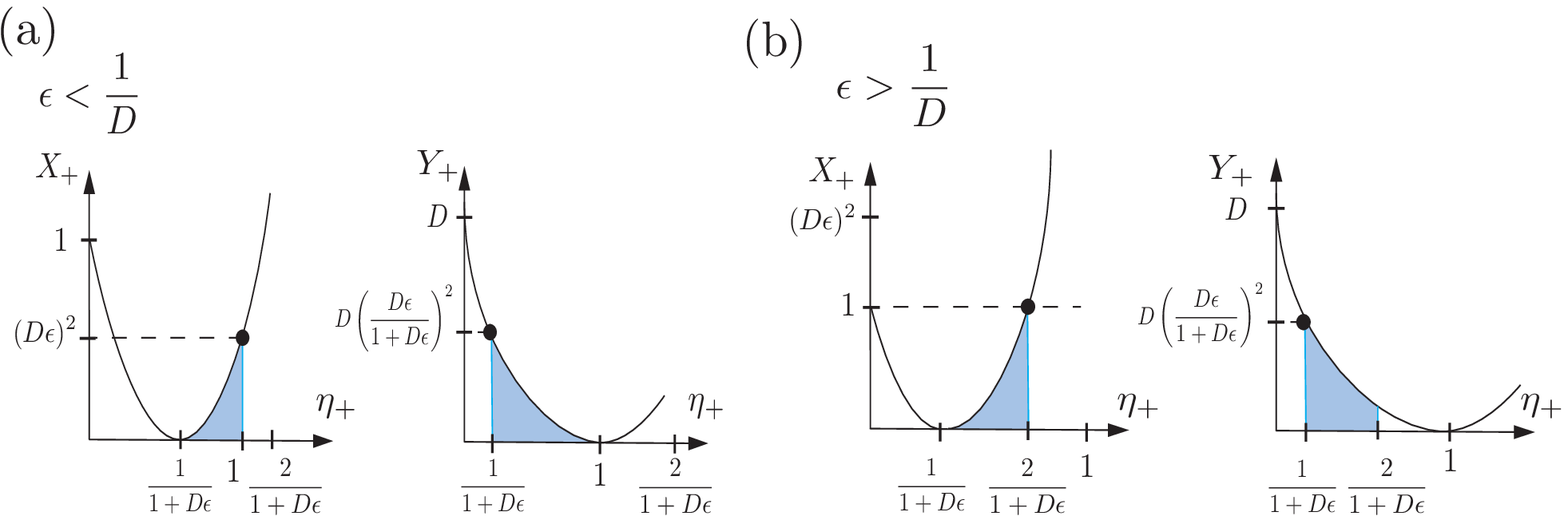}
      \caption{Sketches of the graph of $X_+$ and $Y_+$ for $D\epsilon < 1$ (a) and for $D\epsilon > 1$ (b). 
      }\label{Fig_X+Y+}
\end{figure*}

As a result, we have  
\begin{align}
P_+ 
\leq & 
{\rm min} \biggl \{ 
\inf\limits_{\eta_+ \in R_+} 2 \exp \left ( - \frac{X_+}{4} \right ) , \inf\limits_{\eta_+ \in R_+}2 \exp \left ( - \frac{Y_+}{4} \right ) \biggr \}, 
\end{align} 
where $X_+ \equiv [ (1 + D \epsilon) \eta_+ -1]^2$, $Y_+ \equiv D(1-\eta_+)^2$, and 
\begin{align}
R_+ : \frac{1}{1+ D \epsilon} < \eta_+ < {\rm min} \left \{ 1, \frac{2}{1 + D \epsilon} \right \} . 
\end{align} 
A simple computation shows that 
${\rm max}(X_+) = (D\epsilon)^2$ for $D\epsilon < 1$, 
and ${\rm max}(X_+) = 1$ for $D\epsilon > 1$ (see Fig.~\ref{Fig_X+Y+}), 
which gives 
\begin{align}
      \inf\limits_{\eta_+ \in R_+} 2 \exp \left ( - \frac{X_+}{4} \right ) = & 2 \exp \left [ - \frac{1}{4} {\rm min} \{ 1, (\epsilon D)^2 \} \right ] . \nonumber
      \end{align} 
On the other hand, ${\rm max}(Y_+) =D (D\epsilon)^2/(1+D\epsilon)^2$, which gives 
\begin{align}
\inf\limits_{\eta_+ \in R_+}2 \exp \left ( - \frac{Y_+}{4} \right ) = & 2 \exp \left [ - \frac{D}{4} \left ( \frac{D \epsilon}{1+D \epsilon }\right )^2 \right ] .
\nonumber
\end{align} 
As a result, a simple computation shows that the probability $P_+$ can be bounded as
\begin{align}
P_+ \leq  2 \exp \left [ - \frac{D}{4} \left ( \frac{D \epsilon}{1+D \epsilon }\right )^2 \right ] . 
\label{eqD19}
\end{align}

We now consider the second term in \eqref{eqP+P-}: $P_- \equiv P \left (   | \langle i | \phi \rangle |^2 \leq A_- \right )$. 
For $1/D < \epsilon $, we have $P \left (   | \langle i | \phi \rangle |^2 \leq A_- < 0 \right )= 0$. 
For $\epsilon < 1/D$. 
we consider 
\begin{align}
P_-
= & 
P ( \xi_{2i-1}^2 + \xi_{2i}^2 \leq | {\boldsymbol \xi} |^2 A_-) 
\\ 
\leq & 
{\rm min} \{ P ( \xi_{2i-1}^2 + \xi_{2i}^2 \leq 2 D \eta_- A_-) ,  P (  2 D \eta_- \leq  | {\boldsymbol \xi} |^2  ) \}, 
\end{align} 
for $\eta_- > 0$. 
The first part $ P ( \xi_{2i-1}^2 + \xi_{2i}^2 \leq 2 D \eta_- A_-) $ can be given by 
\begin{align}
& 
P ( \xi_{2i-1}^2 + \xi_{2i}^2 \leq 2 D \eta_- A_-) 
\\ 
= & P \left ( \frac{\xi_{2i-1}^2 + \xi_{2i}^2}{2}  -1 \leq  D \eta_- A_- -1 \right )
\\ 
\leq & P \left ( \left | \frac{\xi_{2i-1}^2 + \xi_{2i}^2  }{2} - 1 \right | \geq  1 - D A_-  \eta_-  \right ). 
\end{align} 
Using concentration inequality (\ref{ConcentrationChi}), we have: 
\begin{align}
P ( \xi_{2i-1}^2 + \xi_{2i}^2 \leq 2 D  A_-\eta_- ) 
\leq & 2 \exp \left [ - \frac{ ( 1- D A_- \eta_-  )^2 }{4} \right ], 
\end{align} 
for $0 < \eta_- < 1/(1-D\epsilon) $. 
By using the two-sided tail bound (\ref{ConcentrationChi}), 
the second part $ P (  2 D \eta_- \leq  | {\boldsymbol \xi} |^2  )$ can be given by 
\begin{align}
P(  2D \eta_-   \leq  | {\boldsymbol \xi} |^2 )  
\leq & 
P\left ( \left  | \frac{1}{2D}  | {\boldsymbol \xi} |^2 -   1 \right |  \geq  \eta_-  -1  \right  )  
\\
\leq & 2 \exp \left [ - \frac{D (\eta_- - 1)^2}{4} \right ], 
\end{align} 
where $1 < \eta_- < 2$. 

\begin{figure*}[tbp]
      \centering
      \includegraphics[width=140mm]{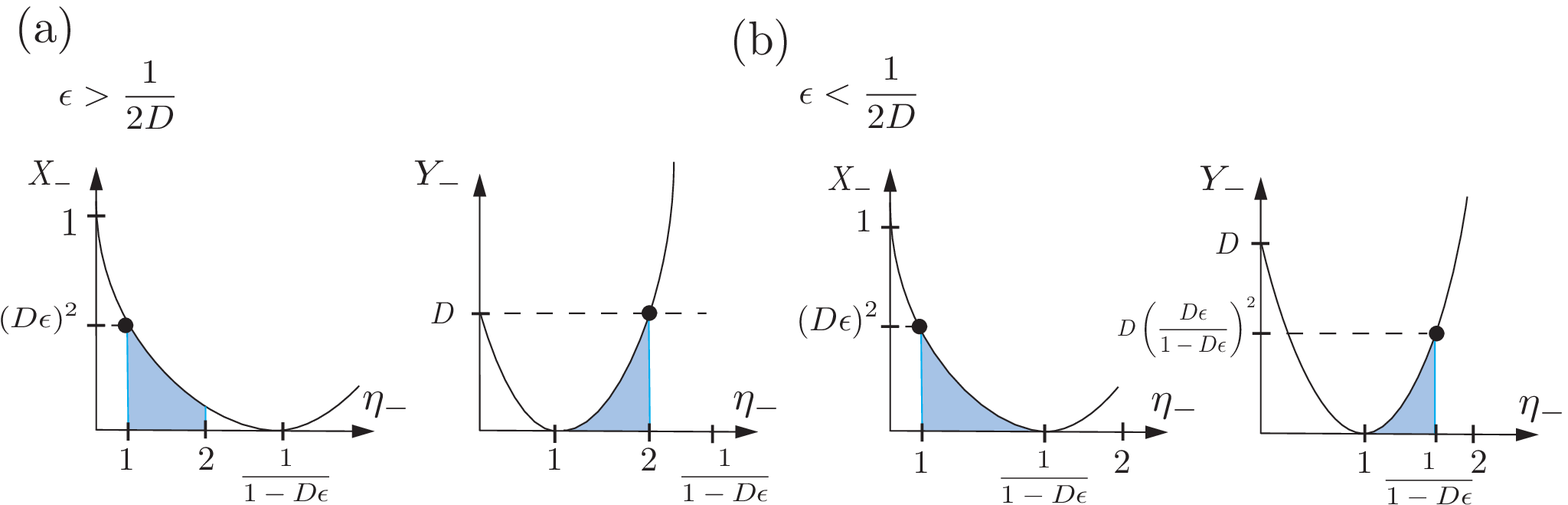}
      \caption{Sketches of the graph of $X_-$ and $Y_-$ for $D\epsilon > 1/2$ (a) and for $D\epsilon < 1/2$ (b). 
      }\label{Fig_X-Y-}
\end{figure*}

As a result, we have  
\begin{align}
P_- 
\leq & 
{\rm min} \biggl \{ 
\inf\limits_{\eta_- \in R_-} 2 \exp \left ( - \frac{X_-}{4} \right ) , \inf\limits_{\eta_- \in R_-}2 \exp \left ( - \frac{Y_-}{4} \right ) \biggr \}, 
\end{align} 
where $X_- \equiv  [ 1 -  \eta_- (1 - D \epsilon) ]^2$, $Y_- \equiv D(\eta_- - 1)^2$, and 
\begin{align}
R_- : 1 < \eta_- < {\rm min} \left \{ 2, \frac{1}{1 - D \epsilon} \right \} . 
\end{align} 
A simple computation shows that 
${\rm max}(X_-) = (D\epsilon)^2$ (see Fig.~\ref{Fig_X-Y-}), which gives 
\begin{align}
      \inf\limits_{\eta_- \in R_-} 2 \exp \left ( - \frac{X_-}{4} \right ) = & 2 \exp \left [ - \frac{ (\epsilon D)^2}{4}  \right ] .
\nonumber
\end{align}
On the other hand, we find ${\rm max}(Y_-) = D$ for $D\epsilon > 1/2$ and ${\rm max}(Y_-) = D(D\epsilon)^2 /(1-D\epsilon)^2 < 1/2$, 
which gives 
\begin{align}
\inf\limits_{\eta_- \in R_-}2 \exp \left ( - \frac{Y_-}{4} \right ) = & 2 \exp \left [ - \frac{D}{4} {\rm min} \left \{ 1, \left ( \frac{D \epsilon}{1-D \epsilon }\right )^2 \right \} \right ] 
\nonumber 
\\
\leq & 2 \exp \left [ - \frac{D}{4} \left ( \frac{D \epsilon}{1+D \epsilon }\right )^2 \right ]. 
\nonumber
\end{align} 
As a result, a simple computation shows that the probability $P_-$ can be bounded as
\begin{align}
P_- \leq  2 \exp \left [ - \frac{D}{4} \left ( \frac{D \epsilon}{1+D \epsilon }\right )^2 \right ] . 
\label{eqD30}
\end{align} 

Using \eqref{eqP+P-}, \eqref{eqD19} and \eqref{eqD30}, 
we finally obtain the concentration inequality 
\begin{align}
P \left (  \left | | \langle i | \phi \rangle |^2  - \frac{1}{D} \right | \geq  \epsilon \right ) \leq 4  \exp \left [ - \frac{D}{4} \left ( \frac{D \epsilon}{1+D \epsilon }\right )^2 \right ] . 
\end{align} 

We here assume that an arbitrary complex vector $| \Psi \rangle$ is generated from a basis $| i \rangle$ by using a unitary matrix $\hat U$, where $| \Psi \rangle = \hat U | i \rangle$. 
Using the same unitary matrix, the state $| \phi \rangle$ is transformed to $|\Psi ' \rangle = \hat U | \phi \rangle = (\Psi_1', \Psi_2', \cdots, \Psi_D')^{\rm T}$, with $\Psi_i' = (\zeta_{2i-1} + i \zeta_{2i})/| {\boldsymbol \zeta} |$ where $| {\boldsymbol \zeta} | = \sum_{i=1}^{2D} \zeta_i^2$. 
If we write $| \phi \rangle = | {\boldsymbol \xi} \rangle / \sqrt{ \langle {\boldsymbol \xi} | {\boldsymbol \xi} \rangle }$ and $| \Psi ' \rangle = | {\boldsymbol \zeta} \rangle / \sqrt{\langle {\boldsymbol \zeta} | {\boldsymbol \zeta}  \rangle }$, we have $| {\boldsymbol \zeta} \rangle = \hat U | {\boldsymbol \xi} \rangle$ and $\sum\limits_{i=1}^{2D} \xi_i^2 = \langle {\boldsymbol \xi} | {\boldsymbol \xi}  \rangle = \langle {\boldsymbol \zeta} | {\boldsymbol \zeta}  \rangle = \sum\limits_{i=1}^{2D} \zeta_i^2$. 
In this case, if ${\boldsymbol \xi}  = (\xi_1, \cdots, \xi_{2D})^{\rm T}$ is generated from the $2D$ normal distribution ${\boldsymbol \xi} \sim {\mathcal N}_{2D} ({\bf 0},{\bf 1})$, 
the $2D$ random vector ${\boldsymbol \zeta}  = (\zeta_1, \cdots, \zeta_{2D})^{\rm T}$ is also uniformly generated from the $2D$ normal distribution ${\boldsymbol \zeta} \sim {\mathcal N}_{2D} ({\bf 0},{\bf 1})$.

To conclude, an arbitrary state $|\Psi \rangle$ and an isotropically random state $|\Psi ' \rangle$ are almost orthogonal for a large dimension Hilbert space, the concentration inequality of which is given in the form of Eq. \eqref{ConcentrateInequality}, or 
\begin{align}
1- 4  \exp \left [ - \frac{D}{4} \left ( \frac{D \epsilon}{1+D \epsilon }\right )^2 \right ] \leq P \left (  \left | | \langle \Psi | \Psi ' \rangle |^2  - \frac{1}{D} \right | <  \epsilon \right ) \leq 1 . 
\end{align}


\bibliography{draft.bib}

\begin{thebibliography}{29}%
\makeatletter
\providecommand \@ifxundefined [1]{%
 \@ifx{#1\undefined}
}%
\providecommand \@ifnum [1]{%
 \ifnum #1\expandafter \@firstoftwo
 \else \expandafter \@secondoftwo
 \fi
}%
\providecommand \@ifx [1]{%
 \ifx #1\expandafter \@firstoftwo
 \else \expandafter \@secondoftwo
 \fi
}%
\providecommand \natexlab [1]{#1}%
\providecommand \enquote  [1]{``#1''}%
\providecommand \bibnamefont  [1]{#1}%
\providecommand \bibfnamefont [1]{#1}%
\providecommand \citenamefont [1]{#1}%
\providecommand \href@noop [0]{\@secondoftwo}%
\providecommand \href [0]{\begingroup \@sanitize@url \@href}%
\providecommand \@href[1]{\@@startlink{#1}\@@href}%
\providecommand \@@href[1]{\endgroup#1\@@endlink}%
\providecommand \@sanitize@url [0]{\catcode `\\12\catcode `\$12\catcode
  `\&12\catcode `\#12\catcode `\^12\catcode `\_12\catcode `\%12\relax}%
\providecommand \@@startlink[1]{}%
\providecommand \@@endlink[0]{}%
\providecommand \url  [0]{\begingroup\@sanitize@url \@url }%
\providecommand \@url [1]{\endgroup\@href {#1}{\urlprefix }}%
\providecommand \urlprefix  [0]{URL }%
\providecommand \Eprint [0]{\href }%
\providecommand \doibase [0]{https://doi.org/}%
\providecommand \selectlanguage [0]{\@gobble}%
\providecommand \bibinfo  [0]{\@secondoftwo}%
\providecommand \bibfield  [0]{\@secondoftwo}%
\providecommand \translation [1]{[#1]}%
\providecommand \BibitemOpen [0]{}%
\providecommand \bibitemStop [0]{}%
\providecommand \bibitemNoStop [0]{.\EOS\space}%
\providecommand \EOS [0]{\spacefactor3000\relax}%
\providecommand \BibitemShut  [1]{\csname bibitem#1\endcsname}%
\let\auto@bib@innerbib\@empty
\bibitem [{\citenamefont {Stauffer}\ and\ \citenamefont
  {Aharony}(1992)}]{stauffer1992}%
  \BibitemOpen
  \bibfield  {author} {\bibinfo {author} {\bibfnamefont {D.}~\bibnamefont
  {Stauffer}}\ and\ \bibinfo {author} {\bibfnamefont {A.}~\bibnamefont
  {Aharony}},\ }\href@noop {} {\emph {\bibinfo {title} {{Introduction To
  Percolation Theory: Second Edition}}}}\ (\bibinfo  {publisher} {(London:
  Taylor \& Francis)},\ \bibinfo {year} {1992})\BibitemShut {NoStop}%
\bibitem [{\citenamefont {Mertens}\ and\ \citenamefont
  {Moore}(2012)}]{Mertens2012}%
  \BibitemOpen
  \bibfield  {author} {\bibinfo {author} {\bibfnamefont {S.}~\bibnamefont
  {Mertens}}\ and\ \bibinfo {author} {\bibfnamefont {C.}~\bibnamefont
  {Moore}},\ }\bibfield  {title} {\bibinfo {title} {{Continuum percolation
  thresholds in two dimensions}},\ }\href@noop {} {\bibfield  {journal}
  {\bibinfo  {journal} {Phys Rev E Stat Nonlin Soft Matter Phys}\ }\textbf
  {\bibinfo {volume} {86}},\ \bibinfo {pages} {061109} (\bibinfo {year}
  {2012})}\BibitemShut {NoStop}%
\bibitem [{\citenamefont {Freedman}(1997)}]{Freedman1997}%
  \BibitemOpen
  \bibfield  {author} {\bibinfo {author} {\bibfnamefont {M.~H.}\ \bibnamefont
  {Freedman}},\ }\bibfield  {title} {\bibinfo {title} {{Percolation on the
  Projective Plane}},\ }\href@noop {} {\bibfield  {journal} {\bibinfo
  {journal} {Mathematical Research Letters}\ }\textbf {\bibinfo {volume} {4}},\
  \bibinfo {pages} {889} (\bibinfo {year} {1997})}\BibitemShut {NoStop}%
\bibitem [{\citenamefont {Borman}\ \emph {et~al.}(2016)\citenamefont {Borman},
  \citenamefont {Grekhov}, \citenamefont {Tronin},\ and\ \citenamefont
  {Tronin}}]{Borman2016}%
  \BibitemOpen
  \bibfield  {author} {\bibinfo {author} {\bibfnamefont {V.~D.}\ \bibnamefont
  {Borman}}, \bibinfo {author} {\bibfnamefont {A.~M.}\ \bibnamefont {Grekhov}},
  \bibinfo {author} {\bibfnamefont {I.~V.}\ \bibnamefont {Tronin}},\ and\
  \bibinfo {author} {\bibfnamefont {V.~N.}\ \bibnamefont {Tronin}},\ }\bibfield
   {title} {\bibinfo {title} {{Percolation threshold of the permeable disks on
  the projective plane}},\ }\href@noop {} {\bibfield  {journal} {\bibinfo
  {journal} {Journal of Physics: Conference Series}\ }\textbf {\bibinfo
  {volume} {751}},\ \bibinfo {pages} {012036} (\bibinfo {year}
  {2016})}\BibitemShut {NoStop}%
\bibitem [{\citenamefont {Kadowaki}\ and\ \citenamefont
  {Nishimori}(1998)}]{Kadowaki1998}%
  \BibitemOpen
  \bibfield  {author} {\bibinfo {author} {\bibfnamefont {T.}~\bibnamefont
  {Kadowaki}}\ and\ \bibinfo {author} {\bibfnamefont {H.}~\bibnamefont
  {Nishimori}},\ }\bibfield  {title} {\bibinfo {title} {Quantum annealing in
  the transverse ising model},\ }\href
  {https://doi.org/10.1103/PhysRevE.58.5355} {\bibfield  {journal} {\bibinfo
  {journal} {Phys. Rev. E}\ }\textbf {\bibinfo {volume} {58}},\ \bibinfo
  {pages} {5355} (\bibinfo {year} {1998})}\BibitemShut {NoStop}%
\bibitem [{\citenamefont {Albash}\ and\ \citenamefont
  {Lidar}(2018)}]{Albash2018}%
  \BibitemOpen
  \bibfield  {author} {\bibinfo {author} {\bibfnamefont {T.}~\bibnamefont
  {Albash}}\ and\ \bibinfo {author} {\bibfnamefont {D.~A.}\ \bibnamefont
  {Lidar}},\ }\bibfield  {title} {\bibinfo {title} {{Adiabatic quantum
  computation}},\ }\href@noop {} {\bibfield  {journal} {\bibinfo  {journal}
  {Rev. Mod. Phys.}\ }\textbf {\bibinfo {volume} {90}},\ \bibinfo {pages}
  {015002} (\bibinfo {year} {2018})}\BibitemShut {NoStop}%
\bibitem [{\citenamefont {Percival}(2008)}]{Percival2008}%
  \BibitemOpen
  \bibfield  {author} {\bibinfo {author} {\bibfnamefont {I.}~\bibnamefont
  {Percival}},\ }\href@noop {} {\emph {\bibinfo {title} {Quantum State
  Diffusion}}}\ (\bibinfo  {publisher} {Cambridge University Press},\ \bibinfo
  {year} {2008})\BibitemShut {NoStop}%
\bibitem [{\citenamefont {Margolus}\ and\ \citenamefont
  {Levitin}(1998)}]{Margolus1998}%
  \BibitemOpen
  \bibfield  {author} {\bibinfo {author} {\bibfnamefont {N.}~\bibnamefont
  {Margolus}}\ and\ \bibinfo {author} {\bibfnamefont {L.~B.}\ \bibnamefont
  {Levitin}},\ }\bibfield  {title} {\bibinfo {title} {{The maximum speed of
  dynamical evolution}},\ }\href@noop {} {\bibfield  {journal} {\bibinfo
  {journal} {Physica D: Nonlinear Phenomena}\ }\textbf {\bibinfo {volume}
  {120}},\ \bibinfo {pages} {188} (\bibinfo {year} {1998})}\BibitemShut
  {NoStop}%
\bibitem [{\citenamefont {Deffner}\ and\ \citenamefont
  {Lutz}(2013)}]{Deffner2013}%
  \BibitemOpen
  \bibfield  {author} {\bibinfo {author} {\bibfnamefont {S.}~\bibnamefont
  {Deffner}}\ and\ \bibinfo {author} {\bibfnamefont {E.}~\bibnamefont {Lutz}},\
  }\bibfield  {title} {\bibinfo {title} {{Energy{\textendash}time uncertainty
  relation for driven quantum systems}},\ }\href@noop {} {\bibfield  {journal}
  {\bibinfo  {journal} {Journal of Physics A: Mathematical and Theoretical}\
  }\textbf {\bibinfo {volume} {46}},\ \bibinfo {pages} {335302} (\bibinfo
  {year} {2013})}\BibitemShut {NoStop}%
\bibitem [{\citenamefont {Okuyama}\ and\ \citenamefont
  {Ohzeki}(2018)}]{Okuyama_2018}%
  \BibitemOpen
  \bibfield  {author} {\bibinfo {author} {\bibfnamefont {M.}~\bibnamefont
  {Okuyama}}\ and\ \bibinfo {author} {\bibfnamefont {M.}~\bibnamefont
  {Ohzeki}},\ }\bibfield  {title} {\bibinfo {title} {Comment on `energy-time
  uncertainty relation for driven quantum systems'},\ }\href
  {https://doi.org/10.1088/1751-8121/aacb90} {\bibfield  {journal} {\bibinfo
  {journal} {Journal of Physics A: Mathematical and Theoretical}\ }\textbf
  {\bibinfo {volume} {51}},\ \bibinfo {pages} {318001} (\bibinfo {year}
  {2018})}\BibitemShut {NoStop}%
\bibitem [{\citenamefont {Wiseman}\ and\ \citenamefont
  {Milburn}(2010)}]{wiseman2010quantum}%
  \BibitemOpen
  \bibfield  {author} {\bibinfo {author} {\bibfnamefont {H.}~\bibnamefont
  {Wiseman}}\ and\ \bibinfo {author} {\bibfnamefont {G.}~\bibnamefont
  {Milburn}},\ }\href {https://books.google.co.jp/books?id=ZNjvHaH8qA4C} {\emph
  {\bibinfo {title} {Quantum Measurement and Control}}}\ (\bibinfo  {publisher}
  {Cambridge University Press},\ \bibinfo {year} {2010})\BibitemShut {NoStop}%
\bibitem [{\citenamefont {Watabe}\ \emph {et~al.}(2022)\citenamefont {Watabe},
  \citenamefont {Serikow}, \citenamefont {Kawabata},\ and\ \citenamefont
  {Zagoskin}}]{Watabe2022}%
  \BibitemOpen
  \bibfield  {author} {\bibinfo {author} {\bibfnamefont {S.}~\bibnamefont
  {Watabe}}, \bibinfo {author} {\bibfnamefont {M.~Z.}\ \bibnamefont {Serikow}},
  \bibinfo {author} {\bibfnamefont {S.}~\bibnamefont {Kawabata}},\ and\
  \bibinfo {author} {\bibfnamefont {A.}~\bibnamefont {Zagoskin}},\ }\bibfield
  {title} {\bibinfo {title} {Efficient criteria of quantumness for a large
  system of qubits},\ }\bibfield  {journal} {\bibinfo  {journal} {Frontiers in
  Physics}\ }\textbf {\bibinfo {volume} {9}},\ \href
  {https://doi.org/10.3389/fphy.2021.773128} {10.3389/fphy.2021.773128}
  (\bibinfo {year} {2022})\BibitemShut {NoStop}%
\bibitem [{\citenamefont {Feynman}(1982)}]{Feynman1982}%
  \BibitemOpen
  \bibfield  {author} {\bibinfo {author} {\bibfnamefont {R.~P.}\ \bibnamefont
  {Feynman}},\ }\bibfield  {title} {\bibinfo {title} {{Simulating physics with
  computers}},\ }\href@noop {} {\bibfield  {journal} {\bibinfo  {journal}
  {International Journal of Theoretical Physics}\ }\textbf {\bibinfo {volume}
  {21}},\ \bibinfo {pages} {467} (\bibinfo {year} {1982})}\BibitemShut
  {NoStop}%
\bibitem [{\citenamefont {Manin}(1980)}]{Manin1980}%
  \BibitemOpen
  \bibfield  {author} {\bibinfo {author} {\bibfnamefont {Y.~I.}\ \bibnamefont
  {Manin}},\ }\href@noop {} {\emph {\bibinfo {title} {Vychislimoe i
  nevychislimoe [Computable and Noncomputable] (in Russian)}}}\ (\bibinfo
  {publisher} {Moscow: Sov. Radio},\ \bibinfo {year} {1980})\BibitemShut
  {NoStop}%
\bibitem [{\citenamefont {de~Touzalin}\ \emph {et~al.}(2016)\citenamefont
  {de~Touzalin}, \citenamefont {Marcus}, \citenamefont {Heijman}, \citenamefont
  {Cirac}, \citenamefont {Murray},\ and\ \citenamefont
  {Calarco}}]{QuantumManifesto2016}%
  \BibitemOpen
  \bibfield  {author} {\bibinfo {author} {\bibfnamefont {A.}~\bibnamefont
  {de~Touzalin}}, \bibinfo {author} {\bibfnamefont {C.}~\bibnamefont {Marcus}},
  \bibinfo {author} {\bibfnamefont {F.}~\bibnamefont {Heijman}}, \bibinfo
  {author} {\bibfnamefont {I.}~\bibnamefont {Cirac}}, \bibinfo {author}
  {\bibfnamefont {R.}~\bibnamefont {Murray}},\ and\ \bibinfo {author}
  {\bibfnamefont {T.}~\bibnamefont {Calarco}},\ }\href
  {http://qurope.eu/manifesto} {\bibinfo {title} {Quantum manifesto. a new era
  of technology}} (\bibinfo {year} {2016})\BibitemShut {NoStop}%
\bibitem [{DWa()}]{DWave2000X}%
  \BibitemOpen
  \href@noop {} {\bibinfo {title} {{Introduction to the D-Wave Quantum
  Hardware}}},\ \bibinfo {howpublished}
  {\url{https://www.dwavesys.com/tutorials/background-reading-series/introduction-d-wave-quantum-hardware}}\BibitemShut
  {NoStop}%
\bibitem [{\citenamefont {Bunyk}\ \emph {et~al.}(2014)\citenamefont {Bunyk},
  \citenamefont {Hoskinson}, \citenamefont {Johnson}, \citenamefont
  {Tolkacheva}, \citenamefont {Altomare}, \citenamefont {Berkley},
  \citenamefont {Harris}, \citenamefont {Hilton}, \citenamefont {Lanting},
  \citenamefont {Przybysz},\ and\ \citenamefont {Whittaker}}]{Bunyk2014}%
  \BibitemOpen
  \bibfield  {author} {\bibinfo {author} {\bibfnamefont {P.~I.}\ \bibnamefont
  {Bunyk}}, \bibinfo {author} {\bibfnamefont {E.~M.}\ \bibnamefont
  {Hoskinson}}, \bibinfo {author} {\bibfnamefont {M.~W.}\ \bibnamefont
  {Johnson}}, \bibinfo {author} {\bibfnamefont {E.}~\bibnamefont {Tolkacheva}},
  \bibinfo {author} {\bibfnamefont {F.}~\bibnamefont {Altomare}}, \bibinfo
  {author} {\bibfnamefont {A.~J.}\ \bibnamefont {Berkley}}, \bibinfo {author}
  {\bibfnamefont {R.}~\bibnamefont {Harris}}, \bibinfo {author} {\bibfnamefont
  {J.~P.}\ \bibnamefont {Hilton}}, \bibinfo {author} {\bibfnamefont
  {T.}~\bibnamefont {Lanting}}, \bibinfo {author} {\bibfnamefont {A.~J.}\
  \bibnamefont {Przybysz}},\ and\ \bibinfo {author} {\bibfnamefont
  {J.}~\bibnamefont {Whittaker}},\ }\bibfield  {title} {\bibinfo {title}
  {{Architectural Considerations in the Design of a Superconducting Quantum
  Annealing Processor}},\ }\href@noop {} {\bibfield  {journal} {\bibinfo
  {journal} {IEEE Transactions on Applied Superconductivity}\ }\textbf
  {\bibinfo {volume} {24}},\ \bibinfo {pages} {1} (\bibinfo {year}
  {2014})}\BibitemShut {NoStop}%
\bibitem [{\citenamefont {Walport}\ and\ \citenamefont
  {Knight}(2016)}]{Walport2016}%
  \BibitemOpen
  \bibfield  {author} {\bibinfo {author} {\bibfnamefont {M.}~\bibnamefont
  {Walport}}\ and\ \bibinfo {author} {\bibfnamefont {P.}~\bibnamefont
  {Knight}},\ }\href
  {https://www.gov.uk/government/publications/quantum-technologies-blackett-review}
  {\bibinfo {title} {The quantum age: technological opportunities}} (\bibinfo
  {year} {2016})\BibitemShut {NoStop}%
\bibitem [{\citenamefont {Arute}\ \emph {et~al.}(2019)\citenamefont {Arute},
  \citenamefont {Arya}, \citenamefont {Babbush}, \citenamefont {Bacon},
  \citenamefont {Bardin}, \citenamefont {Barends}, \citenamefont {Biswas},
  \citenamefont {Boixo}, \citenamefont {Brandao}, \citenamefont {Buell},
  \citenamefont {Burkett}, \citenamefont {Chen}, \citenamefont {Chen},
  \citenamefont {Chiaro}, \citenamefont {Collins}, \citenamefont {Courtney},
  \citenamefont {Dunsworth}, \citenamefont {Farhi}, \citenamefont {Foxen},
  \citenamefont {Fowler}, \citenamefont {Gidney}, \citenamefont {Giustina},
  \citenamefont {Graff}, \citenamefont {Guerin}, \citenamefont {Habegger},
  \citenamefont {Harrigan}, \citenamefont {Hartmann}, \citenamefont {Ho},
  \citenamefont {Hoffmann}, \citenamefont {Huang}, \citenamefont {Humble},
  \citenamefont {Isakov}, \citenamefont {Jeffrey}, \citenamefont {Jiang},
  \citenamefont {Kafri}, \citenamefont {Kechedzhi}, \citenamefont {Kelly},
  \citenamefont {Klimov}, \citenamefont {Knysh}, \citenamefont {Korotkov},
  \citenamefont {Kostritsa}, \citenamefont {Landhuis}, \citenamefont
  {Lindmark}, \citenamefont {Lucero}, \citenamefont {Lyakh}, \citenamefont
  {Mandr{\`a}}, \citenamefont {McClean}, \citenamefont {McEwen}, \citenamefont
  {Megrant}, \citenamefont {Mi}, \citenamefont {Michielsen}, \citenamefont
  {Mohseni}, \citenamefont {Mutus}, \citenamefont {Naaman}, \citenamefont
  {Neeley}, \citenamefont {Neill}, \citenamefont {Niu}, \citenamefont {Ostby},
  \citenamefont {Petukhov}, \citenamefont {Platt}, \citenamefont {Quintana},
  \citenamefont {Rieffel}, \citenamefont {Roushan}, \citenamefont {Rubin},
  \citenamefont {Sank}, \citenamefont {Satzinger}, \citenamefont {Smelyanskiy},
  \citenamefont {Sung}, \citenamefont {Trevithick}, \citenamefont
  {Vainsencher}, \citenamefont {Villalonga}, \citenamefont {White},
  \citenamefont {Yao}, \citenamefont {Yeh}, \citenamefont {Zalcman},
  \citenamefont {Neven},\ and\ \citenamefont {Martinis}}]{Arute2019}%
  \BibitemOpen
  \bibfield  {author} {\bibinfo {author} {\bibfnamefont {F.}~\bibnamefont
  {Arute}}, \bibinfo {author} {\bibfnamefont {K.}~\bibnamefont {Arya}},
  \bibinfo {author} {\bibfnamefont {R.}~\bibnamefont {Babbush}}, \bibinfo
  {author} {\bibfnamefont {D.}~\bibnamefont {Bacon}}, \bibinfo {author}
  {\bibfnamefont {J.~C.}\ \bibnamefont {Bardin}}, \bibinfo {author}
  {\bibfnamefont {R.}~\bibnamefont {Barends}}, \bibinfo {author} {\bibfnamefont
  {R.}~\bibnamefont {Biswas}}, \bibinfo {author} {\bibfnamefont
  {S.}~\bibnamefont {Boixo}}, \bibinfo {author} {\bibfnamefont {F.~G. S.~L.}\
  \bibnamefont {Brandao}}, \bibinfo {author} {\bibfnamefont {D.~A.}\
  \bibnamefont {Buell}}, \bibinfo {author} {\bibfnamefont {B.}~\bibnamefont
  {Burkett}}, \bibinfo {author} {\bibfnamefont {Y.}~\bibnamefont {Chen}},
  \bibinfo {author} {\bibfnamefont {Z.}~\bibnamefont {Chen}}, \bibinfo {author}
  {\bibfnamefont {B.}~\bibnamefont {Chiaro}}, \bibinfo {author} {\bibfnamefont
  {R.}~\bibnamefont {Collins}}, \bibinfo {author} {\bibfnamefont
  {W.}~\bibnamefont {Courtney}}, \bibinfo {author} {\bibfnamefont
  {A.}~\bibnamefont {Dunsworth}}, \bibinfo {author} {\bibfnamefont
  {E.}~\bibnamefont {Farhi}}, \bibinfo {author} {\bibfnamefont
  {B.}~\bibnamefont {Foxen}}, \bibinfo {author} {\bibfnamefont
  {A.}~\bibnamefont {Fowler}}, \bibinfo {author} {\bibfnamefont
  {C.}~\bibnamefont {Gidney}}, \bibinfo {author} {\bibfnamefont
  {M.}~\bibnamefont {Giustina}}, \bibinfo {author} {\bibfnamefont
  {R.}~\bibnamefont {Graff}}, \bibinfo {author} {\bibfnamefont
  {K.}~\bibnamefont {Guerin}}, \bibinfo {author} {\bibfnamefont
  {S.}~\bibnamefont {Habegger}}, \bibinfo {author} {\bibfnamefont {M.~P.}\
  \bibnamefont {Harrigan}}, \bibinfo {author} {\bibfnamefont {M.~J.}\
  \bibnamefont {Hartmann}}, \bibinfo {author} {\bibfnamefont {A.}~\bibnamefont
  {Ho}}, \bibinfo {author} {\bibfnamefont {M.}~\bibnamefont {Hoffmann}},
  \bibinfo {author} {\bibfnamefont {T.}~\bibnamefont {Huang}}, \bibinfo
  {author} {\bibfnamefont {T.~S.}\ \bibnamefont {Humble}}, \bibinfo {author}
  {\bibfnamefont {S.~V.}\ \bibnamefont {Isakov}}, \bibinfo {author}
  {\bibfnamefont {E.}~\bibnamefont {Jeffrey}}, \bibinfo {author} {\bibfnamefont
  {Z.}~\bibnamefont {Jiang}}, \bibinfo {author} {\bibfnamefont
  {D.}~\bibnamefont {Kafri}}, \bibinfo {author} {\bibfnamefont
  {K.}~\bibnamefont {Kechedzhi}}, \bibinfo {author} {\bibfnamefont
  {J.}~\bibnamefont {Kelly}}, \bibinfo {author} {\bibfnamefont {P.~V.}\
  \bibnamefont {Klimov}}, \bibinfo {author} {\bibfnamefont {S.}~\bibnamefont
  {Knysh}}, \bibinfo {author} {\bibfnamefont {A.}~\bibnamefont {Korotkov}},
  \bibinfo {author} {\bibfnamefont {F.}~\bibnamefont {Kostritsa}}, \bibinfo
  {author} {\bibfnamefont {D.}~\bibnamefont {Landhuis}}, \bibinfo {author}
  {\bibfnamefont {M.}~\bibnamefont {Lindmark}}, \bibinfo {author}
  {\bibfnamefont {E.}~\bibnamefont {Lucero}}, \bibinfo {author} {\bibfnamefont
  {D.}~\bibnamefont {Lyakh}}, \bibinfo {author} {\bibfnamefont
  {S.}~\bibnamefont {Mandr{\`a}}}, \bibinfo {author} {\bibfnamefont {J.~R.}\
  \bibnamefont {McClean}}, \bibinfo {author} {\bibfnamefont {M.}~\bibnamefont
  {McEwen}}, \bibinfo {author} {\bibfnamefont {A.}~\bibnamefont {Megrant}},
  \bibinfo {author} {\bibfnamefont {X.}~\bibnamefont {Mi}}, \bibinfo {author}
  {\bibfnamefont {K.}~\bibnamefont {Michielsen}}, \bibinfo {author}
  {\bibfnamefont {M.}~\bibnamefont {Mohseni}}, \bibinfo {author} {\bibfnamefont
  {J.}~\bibnamefont {Mutus}}, \bibinfo {author} {\bibfnamefont
  {O.}~\bibnamefont {Naaman}}, \bibinfo {author} {\bibfnamefont
  {M.}~\bibnamefont {Neeley}}, \bibinfo {author} {\bibfnamefont
  {C.}~\bibnamefont {Neill}}, \bibinfo {author} {\bibfnamefont {M.~Y.}\
  \bibnamefont {Niu}}, \bibinfo {author} {\bibfnamefont {E.}~\bibnamefont
  {Ostby}}, \bibinfo {author} {\bibfnamefont {A.}~\bibnamefont {Petukhov}},
  \bibinfo {author} {\bibfnamefont {J.~C.}\ \bibnamefont {Platt}}, \bibinfo
  {author} {\bibfnamefont {C.}~\bibnamefont {Quintana}}, \bibinfo {author}
  {\bibfnamefont {E.~G.}\ \bibnamefont {Rieffel}}, \bibinfo {author}
  {\bibfnamefont {P.}~\bibnamefont {Roushan}}, \bibinfo {author} {\bibfnamefont
  {N.~C.}\ \bibnamefont {Rubin}}, \bibinfo {author} {\bibfnamefont
  {D.}~\bibnamefont {Sank}}, \bibinfo {author} {\bibfnamefont {K.~J.}\
  \bibnamefont {Satzinger}}, \bibinfo {author} {\bibfnamefont {V.}~\bibnamefont
  {Smelyanskiy}}, \bibinfo {author} {\bibfnamefont {K.~J.}\ \bibnamefont
  {Sung}}, \bibinfo {author} {\bibfnamefont {M.~D.}\ \bibnamefont
  {Trevithick}}, \bibinfo {author} {\bibfnamefont {A.}~\bibnamefont
  {Vainsencher}}, \bibinfo {author} {\bibfnamefont {B.}~\bibnamefont
  {Villalonga}}, \bibinfo {author} {\bibfnamefont {T.}~\bibnamefont {White}},
  \bibinfo {author} {\bibfnamefont {Z.~J.}\ \bibnamefont {Yao}}, \bibinfo
  {author} {\bibfnamefont {P.}~\bibnamefont {Yeh}}, \bibinfo {author}
  {\bibfnamefont {A.}~\bibnamefont {Zalcman}}, \bibinfo {author} {\bibfnamefont
  {H.}~\bibnamefont {Neven}},\ and\ \bibinfo {author} {\bibfnamefont {J.~M.}\
  \bibnamefont {Martinis}},\ }\bibfield  {title} {\bibinfo {title} {Quantum
  supremacy using a programmable superconducting processor},\ }\href
  {https://doi.org/10.1038/s41586-019-1666-5} {\bibfield  {journal} {\bibinfo
  {journal} {Nature}\ }\textbf {\bibinfo {volume} {574}},\ \bibinfo {pages}
  {505} (\bibinfo {year} {2019})}\BibitemShut {NoStop}%
\bibitem [{IBM()}]{IBM127}%
  \BibitemOpen
  \href@noop {} {\bibinfo {title} {{IBM Quantum breaks the 100^^e2^^80^^91qubit
  processor barrier}}},\ \bibinfo {howpublished}
  {\url{https://research.ibm.com/blog/127-qubit-quantum-processor-eagle}}\BibitemShut
  {NoStop}%
\bibitem [{\citenamefont {Gong}\ \emph {et~al.}(2021)\citenamefont {Gong},
  \citenamefont {Wang}, \citenamefont {Zha}, \citenamefont {Chen},
  \citenamefont {Huang}, \citenamefont {Wu}, \citenamefont {Zhu}, \citenamefont
  {Zhao}, \citenamefont {Li}, \citenamefont {Guo}, \citenamefont {Qian},
  \citenamefont {Ye}, \citenamefont {Chen}, \citenamefont {Ying}, \citenamefont
  {Yu}, \citenamefont {Fan}, \citenamefont {Wu}, \citenamefont {Su},
  \citenamefont {Deng}, \citenamefont {Rong}, \citenamefont {Zhang},
  \citenamefont {Cao}, \citenamefont {Lin}, \citenamefont {Xu}, \citenamefont
  {Sun}, \citenamefont {Guo}, \citenamefont {Li}, \citenamefont {Liang},
  \citenamefont {Bastidas}, \citenamefont {Nemoto}, \citenamefont {Munro},
  \citenamefont {Huo}, \citenamefont {Lu}, \citenamefont {Peng}, \citenamefont
  {Zhu},\ and\ \citenamefont {Pan}}]{Gong2021}%
  \BibitemOpen
  \bibfield  {author} {\bibinfo {author} {\bibfnamefont {M.}~\bibnamefont
  {Gong}}, \bibinfo {author} {\bibfnamefont {S.}~\bibnamefont {Wang}}, \bibinfo
  {author} {\bibfnamefont {C.}~\bibnamefont {Zha}}, \bibinfo {author}
  {\bibfnamefont {M.-C.}\ \bibnamefont {Chen}}, \bibinfo {author}
  {\bibfnamefont {H.-L.}\ \bibnamefont {Huang}}, \bibinfo {author}
  {\bibfnamefont {Y.}~\bibnamefont {Wu}}, \bibinfo {author} {\bibfnamefont
  {Q.}~\bibnamefont {Zhu}}, \bibinfo {author} {\bibfnamefont {Y.}~\bibnamefont
  {Zhao}}, \bibinfo {author} {\bibfnamefont {S.}~\bibnamefont {Li}}, \bibinfo
  {author} {\bibfnamefont {S.}~\bibnamefont {Guo}}, \bibinfo {author}
  {\bibfnamefont {H.}~\bibnamefont {Qian}}, \bibinfo {author} {\bibfnamefont
  {Y.}~\bibnamefont {Ye}}, \bibinfo {author} {\bibfnamefont {F.}~\bibnamefont
  {Chen}}, \bibinfo {author} {\bibfnamefont {C.}~\bibnamefont {Ying}}, \bibinfo
  {author} {\bibfnamefont {J.}~\bibnamefont {Yu}}, \bibinfo {author}
  {\bibfnamefont {D.}~\bibnamefont {Fan}}, \bibinfo {author} {\bibfnamefont
  {D.}~\bibnamefont {Wu}}, \bibinfo {author} {\bibfnamefont {H.}~\bibnamefont
  {Su}}, \bibinfo {author} {\bibfnamefont {H.}~\bibnamefont {Deng}}, \bibinfo
  {author} {\bibfnamefont {H.}~\bibnamefont {Rong}}, \bibinfo {author}
  {\bibfnamefont {K.}~\bibnamefont {Zhang}}, \bibinfo {author} {\bibfnamefont
  {S.}~\bibnamefont {Cao}}, \bibinfo {author} {\bibfnamefont {J.}~\bibnamefont
  {Lin}}, \bibinfo {author} {\bibfnamefont {Y.}~\bibnamefont {Xu}}, \bibinfo
  {author} {\bibfnamefont {L.}~\bibnamefont {Sun}}, \bibinfo {author}
  {\bibfnamefont {C.}~\bibnamefont {Guo}}, \bibinfo {author} {\bibfnamefont
  {N.}~\bibnamefont {Li}}, \bibinfo {author} {\bibfnamefont {F.}~\bibnamefont
  {Liang}}, \bibinfo {author} {\bibfnamefont {V.~M.}\ \bibnamefont {Bastidas}},
  \bibinfo {author} {\bibfnamefont {K.}~\bibnamefont {Nemoto}}, \bibinfo
  {author} {\bibfnamefont {W.~J.}\ \bibnamefont {Munro}}, \bibinfo {author}
  {\bibfnamefont {Y.-H.}\ \bibnamefont {Huo}}, \bibinfo {author} {\bibfnamefont
  {C.-Y.}\ \bibnamefont {Lu}}, \bibinfo {author} {\bibfnamefont {C.-Z.}\
  \bibnamefont {Peng}}, \bibinfo {author} {\bibfnamefont {X.}~\bibnamefont
  {Zhu}},\ and\ \bibinfo {author} {\bibfnamefont {J.-W.}\ \bibnamefont {Pan}},\
  }\bibfield  {title} {\bibinfo {title} {Quantum walks on a programmable
  two-dimensional 62-qubit superconducting processor},\ }\href
  {https://doi.org/10.1126/science.abg7812} {\bibfield  {journal} {\bibinfo
  {journal} {Science}\ }\textbf {\bibinfo {volume} {372}},\ \bibinfo {pages}
  {948} (\bibinfo {year} {2021})},\ \Eprint
  {https://arxiv.org/abs/https://www.science.org/doi/pdf/10.1126/science.abg7812}
  {https://www.science.org/doi/pdf/10.1126/science.abg7812} \BibitemShut
  {NoStop}%
\bibitem [{\citenamefont {Ebadi}\ \emph {et~al.}(2021)\citenamefont {Ebadi},
  \citenamefont {Wang}, \citenamefont {Levine}, \citenamefont {Keesling},
  \citenamefont {Semeghini}, \citenamefont {Omran}, \citenamefont {Bluvstein},
  \citenamefont {Samajdar}, \citenamefont {Pichler}, \citenamefont {Ho},
  \citenamefont {Choi}, \citenamefont {Sachdev}, \citenamefont {Greiner},
  \citenamefont {Vuleti{\'{c}}},\ and\ \citenamefont {Lukin}}]{Ebadi2021}%
  \BibitemOpen
  \bibfield  {author} {\bibinfo {author} {\bibfnamefont {S.}~\bibnamefont
  {Ebadi}}, \bibinfo {author} {\bibfnamefont {T.~T.}\ \bibnamefont {Wang}},
  \bibinfo {author} {\bibfnamefont {H.}~\bibnamefont {Levine}}, \bibinfo
  {author} {\bibfnamefont {A.}~\bibnamefont {Keesling}}, \bibinfo {author}
  {\bibfnamefont {G.}~\bibnamefont {Semeghini}}, \bibinfo {author}
  {\bibfnamefont {A.}~\bibnamefont {Omran}}, \bibinfo {author} {\bibfnamefont
  {D.}~\bibnamefont {Bluvstein}}, \bibinfo {author} {\bibfnamefont
  {R.}~\bibnamefont {Samajdar}}, \bibinfo {author} {\bibfnamefont
  {H.}~\bibnamefont {Pichler}}, \bibinfo {author} {\bibfnamefont {W.~W.}\
  \bibnamefont {Ho}}, \bibinfo {author} {\bibfnamefont {S.}~\bibnamefont
  {Choi}}, \bibinfo {author} {\bibfnamefont {S.}~\bibnamefont {Sachdev}},
  \bibinfo {author} {\bibfnamefont {M.}~\bibnamefont {Greiner}}, \bibinfo
  {author} {\bibfnamefont {V.}~\bibnamefont {Vuleti{\'{c}}}},\ and\ \bibinfo
  {author} {\bibfnamefont {M.~D.}\ \bibnamefont {Lukin}},\ }\bibfield  {title}
  {\bibinfo {title} {Quantum phases of matter on a 256-atom programmable
  quantum simulator},\ }\href {https://doi.org/10.1038/s41586-021-03582-4}
  {\bibfield  {journal} {\bibinfo  {journal} {Nature}\ }\textbf {\bibinfo
  {volume} {595}},\ \bibinfo {pages} {227} (\bibinfo {year}
  {2021})}\BibitemShut {NoStop}%
\bibitem [{\citenamefont {Madsen}\ \emph {et~al.}(2022)\citenamefont {Madsen},
  \citenamefont {Laudenbach}, \citenamefont {Askarani}, \citenamefont
  {Rortais}, \citenamefont {Vincent}, \citenamefont {Bulmer}, \citenamefont
  {Miatto}, \citenamefont {Neuhaus}, \citenamefont {Helt}, \citenamefont
  {Collins}, \citenamefont {Lita}, \citenamefont {Gerrits}, \citenamefont
  {Nam}, \citenamefont {Vaidya}, \citenamefont {Menotti}, \citenamefont
  {Dhand}, \citenamefont {Vernon}, \citenamefont {Quesada},\ and\ \citenamefont
  {Lavoie}}]{Madsen2022}%
  \BibitemOpen
  \bibfield  {author} {\bibinfo {author} {\bibfnamefont {L.~S.}\ \bibnamefont
  {Madsen}}, \bibinfo {author} {\bibfnamefont {F.}~\bibnamefont {Laudenbach}},
  \bibinfo {author} {\bibfnamefont {M.~F.}\ \bibnamefont {Askarani}}, \bibinfo
  {author} {\bibfnamefont {F.}~\bibnamefont {Rortais}}, \bibinfo {author}
  {\bibfnamefont {T.}~\bibnamefont {Vincent}}, \bibinfo {author} {\bibfnamefont
  {J.~F.~F.}\ \bibnamefont {Bulmer}}, \bibinfo {author} {\bibfnamefont {F.~M.}\
  \bibnamefont {Miatto}}, \bibinfo {author} {\bibfnamefont {L.}~\bibnamefont
  {Neuhaus}}, \bibinfo {author} {\bibfnamefont {L.~G.}\ \bibnamefont {Helt}},
  \bibinfo {author} {\bibfnamefont {M.~J.}\ \bibnamefont {Collins}}, \bibinfo
  {author} {\bibfnamefont {A.~E.}\ \bibnamefont {Lita}}, \bibinfo {author}
  {\bibfnamefont {T.}~\bibnamefont {Gerrits}}, \bibinfo {author} {\bibfnamefont
  {S.~W.}\ \bibnamefont {Nam}}, \bibinfo {author} {\bibfnamefont {V.~D.}\
  \bibnamefont {Vaidya}}, \bibinfo {author} {\bibfnamefont {M.}~\bibnamefont
  {Menotti}}, \bibinfo {author} {\bibfnamefont {I.}~\bibnamefont {Dhand}},
  \bibinfo {author} {\bibfnamefont {Z.}~\bibnamefont {Vernon}}, \bibinfo
  {author} {\bibfnamefont {N.}~\bibnamefont {Quesada}},\ and\ \bibinfo {author}
  {\bibfnamefont {J.}~\bibnamefont {Lavoie}},\ }\bibfield  {title} {\bibinfo
  {title} {Quantum computational advantage with a programmable photonic
  processor},\ }\href {https://doi.org/10.1038/s41586-022-04725-x} {\bibfield
  {journal} {\bibinfo  {journal} {Nature}\ }\textbf {\bibinfo {volume} {606}},\
  \bibinfo {pages} {75} (\bibinfo {year} {2022})}\BibitemShut {NoStop}%
\bibitem [{\citenamefont {Bravyi}\ \emph {et~al.}(2022)\citenamefont {Bravyi},
  \citenamefont {Dial}, \citenamefont {Gambetta}, \citenamefont {Gil},\ and\
  \citenamefont {Nazario}}]{arxiv.2209.06841}%
  \BibitemOpen
  \bibfield  {author} {\bibinfo {author} {\bibfnamefont {S.}~\bibnamefont
  {Bravyi}}, \bibinfo {author} {\bibfnamefont {O.}~\bibnamefont {Dial}},
  \bibinfo {author} {\bibfnamefont {J.~M.}\ \bibnamefont {Gambetta}}, \bibinfo
  {author} {\bibfnamefont {D.}~\bibnamefont {Gil}},\ and\ \bibinfo {author}
  {\bibfnamefont {Z.}~\bibnamefont {Nazario}},\ }\href
  {https://doi.org/10.48550/ARXIV.2209.06841} {\bibinfo {title} {The future of
  quantum computing with superconducting qubits}} (\bibinfo {year}
  {2022})\BibitemShut {NoStop}%
\bibitem [{\citenamefont {Stauffer}\ and\ \citenamefont
  {Aharony}(2018)}]{stauffer2018introduction}%
  \BibitemOpen
  \bibfield  {author} {\bibinfo {author} {\bibfnamefont {D.}~\bibnamefont
  {Stauffer}}\ and\ \bibinfo {author} {\bibfnamefont {A.}~\bibnamefont
  {Aharony}},\ }\href {https://books.google.co.jp/books?id=E0ZZDwAAQBAJ} {\emph
  {\bibinfo {title} {Introduction To Percolation Theory: Second Edition}}}\
  (\bibinfo  {publisher} {CRC Press},\ \bibinfo {year} {2018})\BibitemShut
  {NoStop}%
\bibitem [{\citenamefont {Brody}\ and\ \citenamefont
  {Hughston}(2001)}]{Brody2001}%
  \BibitemOpen
  \bibfield  {author} {\bibinfo {author} {\bibfnamefont {D.~C.}\ \bibnamefont
  {Brody}}\ and\ \bibinfo {author} {\bibfnamefont {L.~P.}\ \bibnamefont
  {Hughston}},\ }\bibfield  {title} {\bibinfo {title} {{Geometric quantum
  mechanics}},\ }\href@noop {} {\bibfield  {journal} {\bibinfo  {journal}
  {Journal of Geometry and Physics}\ }\textbf {\bibinfo {volume} {38}},\
  \bibinfo {pages} {19} (\bibinfo {year} {2001})}\BibitemShut {NoStop}%
\bibitem [{\citenamefont {Vershynin}(2018)}]{vershynin_2018}%
  \BibitemOpen
  \bibfield  {author} {\bibinfo {author} {\bibfnamefont {R.}~\bibnamefont
  {Vershynin}},\ }\href {https://doi.org/10.1017/9781108231596} {\emph
  {\bibinfo {title} {High-Dimensional Probability: An Introduction with
  Applications in Data Science}}},\ Cambridge Series in Statistical and
  Probabilistic Mathematics\ (\bibinfo  {publisher} {Cambridge University
  Press},\ \bibinfo {year} {2018})\BibitemShut {NoStop}%
\bibitem [{\citenamefont {Hall}\ \emph {et~al.}(2005)\citenamefont {Hall},
  \citenamefont {Marron},\ and\ \citenamefont {Neeman}}]{Hall2005}%
  \BibitemOpen
  \bibfield  {author} {\bibinfo {author} {\bibfnamefont {P.}~\bibnamefont
  {Hall}}, \bibinfo {author} {\bibfnamefont {J.~S.}\ \bibnamefont {Marron}},\
  and\ \bibinfo {author} {\bibfnamefont {A.}~\bibnamefont {Neeman}},\
  }\bibfield  {title} {\bibinfo {title} {Geometric representation of high
  dimension, low sample size data},\ }\href
  {https://doi.org/10.1111/j.1467-9868.2005.00510.x} {\bibfield  {journal}
  {\bibinfo  {journal} {Journal of the Royal Statistical Society: Series B
  (Statistical Methodology)}\ }\textbf {\bibinfo {volume} {67}},\ \bibinfo
  {pages} {427} (\bibinfo {year} {2005})}\BibitemShut {NoStop}%
\bibitem [{\citenamefont {Wainwright}(2019)}]{wainwright_2019}%
  \BibitemOpen
  \bibfield  {author} {\bibinfo {author} {\bibfnamefont {M.~J.}\ \bibnamefont
  {Wainwright}},\ }\href {https://doi.org/10.1017/9781108627771} {\emph
  {\bibinfo {title} {High-Dimensional Statistics: A Non-Asymptotic
  Viewpoint}}},\ Cambridge Series in Statistical and Probabilistic Mathematics\
  (\bibinfo  {publisher} {Cambridge University Press},\ \bibinfo {year}
  {2019})\BibitemShut {NoStop}%
\end{thebibliography}%

\end{document}